\documentclass[conference,letterpaper]{IEEEtran}
\usepackage{cite}

\ifCLASSINFOpdf
  \usepackage[pdftex]{graphicx}
  \graphicspath{{plots_graphics/}}
  \DeclareGraphicsExtensions{.eps}
\else
  \usepackage[dvips]{graphicx}
  \graphicspath{{plots_graphics/}}
	\DeclareGraphicsExtensions{.eps}
\fi
\usepackage[cmex10]{amsmath}
\usepackage{algorithmic}

\usepackage{array}

\usepackage{mdwmath}
\usepackage{mdwtab}

\usepackage{eqparbox}

\usepackage[tight,footnotesize]{subfigure}

\usepackage[caption=false,font=footnotesize]{subfig}
\usepackage{fixltx2e}

\usepackage{stfloats}

\usepackage{url}

\begin{document}
%
\title{Hypermodular Self-Assembling Space Solar Power - \\ Design Option for Mid-Term GEO Utility-Scale Power Plants}

\author{
\IEEEauthorblockN{Martin Leitgab, PhD}
\IEEEauthorblockA{
Physics Department\\
University of Illinois at Urbana-Champaign\\
1110 West Green Street, Urbana, IL 61801\\
Email: Martin.Leitgab@gmail.com
}
}


%


\maketitle

\begin{abstract}
This paper presents a design for scaleable space solar power systems based on free-flying reflectors and module self-assembly. Lower system cost of utility-scale space solar power is achieved by design independence of yet-to-be-built in-space assembly or transportation infrastructure. Using current and expected near-term technology, this study describe a design for mid-term utility-scale power plants in geosynchronous orbits. High-level economic considerations in the context of current and expected future launch costs are given as well.
\end{abstract}


%
\IEEEpeerreviewmaketitle

\section{Introduction}
\label{sec:intro}

\subsection{Space Solar Power}

The concept of space solar power (SSP) bases on conversion of solar energy into electricity on orbit and transmission of the collected energy to Earth through wireless power transmission. The concept was first proposed in Ref.~\cite{glaser} and continuously attracted interest from researchers and government agencies. Several system-level studies established technological feasibility of SSP (e.g., the $2011$ IAA study, Ref.~\cite{iaastudy2011}) and potential economic viability (e.g., Ref.~\cite{mankinsniac}). Multiple subsystem technology verifications and demonstrations have been carried out (for an overview see Refs.~\cite{iaastudy2011} and~\cite{leopoldpaper}), and first prototype integration studies including space environment simulation were performed at the Naval Research Laboratory, USA~\cite{jaffepaper}. Space solar power represents a candidate large-scale renewable energy source in the context of rising global energy demand, which is expected to increase by $30\%$ until $2035$~\cite{ieareport}. Compared to energy production from non-renewable energies, e.g. fossil fuels, electricity from SSP causes lower emissions of greenhouse gases which are connected to man-made climate change~\cite{ipcc4threportwg1}. 

\subsection{Hypermodular Self-Assembling Space Solar Power (HSA-SSP)}

This paper in parts bases on research work reported in an entry to the $2^{\textnormal{nd}}$ Space Solar Power International Student and Young Professional Design Competition of the Space Generation Advisory Council/International Astronautical Federation which was selected as the winning contribution and which will be published in the International Astronautical Congress $2013$ proceedings~\cite{ownIACpaper}. While the eventual SSP constellation architecture is entirely different, similarities between Ref.~\cite{ownIACpaper} and the design presented in this paper will be explicitly pointed out. 

The goal of this paper is the same as in Ref.~\cite{ownIACpaper}: to contribute to enabling mid-term SSP deployment. To achieve this, the latest experimental SSP system integration results are combined with aspects of previous SSP system architecture work to formulate a scalable SSP design, under special attention to independence of in-space assembly and transportation infrastructure. The design avoids such auxiliary systems since their development is likely to delay or even prevent realization of utility-scale SSP as a whole due to the larger amount of required resources to achieve first power. 

The SSP design presented in this paper similarly to Ref.~\cite{ownIACpaper} also bases heavily on hypermodularity to reduce total system cost through mass production and increase system reliability due to fewer points of failure (as recently shown in e.g. Ref.~\cite{mankinsniac}). Thus the design is named Hypermodular Self-Assembling Space Solar Power (HSA-SSP). Another feature of SSP designs first introduced in Ref.~\cite{ownIACpaper} is the limitation of SSP satellite size to the capacities of near-term launch vehicles, such as SpaceX' Falcon Heavy launcher~\cite{spacex}, to ensure realistic self-assembly. This requirement allows insect-like formation of large structures or constellations by identical, but yet independently fully functional SPS, similar but different to the ideas presented in Ref.~\cite{mankinsniac}. In addition, the concept of such fragmentation of large structures into fully functional units enables the begin of power production already after the delivery of a certain smaller threshold amount of units, which allows revenue income from early on in the deployment time frame and shortens the time to amortization of invested cost. 

Similarly to the analysis performed in Ref.~\cite{ownIACpaper}, the design of HSA-SSP utilizes photovoltaic cells for electricity generation and microwave power transmission at $2.45$~GHz using solid state power amplifiers for simplicity. 


\section{Specifications, Sizing and Power Scales of HSA-SPS System Layout}

Along the descriptions in Ref.~\cite{ownIACpaper}, each HSA-SPS consists of a main planar platform composed of several hexagonally-shaped 'sandwich' structures where the power generation and power transmission surfaces are located close to each other in a parallel, sandwich-like layering, similar in design to the 'hexbus' sandwich structures proposed in Ref.~\cite{mankinsniac}. The main platform also contains all other systems necessary to operate the satellite in space, such as thermal management systems, guidance, navigation and control, command and data systems, and communication systems. 

The HSA-SPS power transmission surface of the sandwich modules is pointed towards earth at all times. In order to achieve continuous illumination of the power generating surface throughout the entire orbit, each HSA-SPS platform is associated with a free-flying associated reflector or concentrator structure. The reflector structure is of sufficient size to allow three-sun concentration on the photovoltaic sandwich surface and can consist of individually controlled movable reflector elements, but will otherwise not be specified in more detail in this paper. The reflectors are not structurally connected to the sandwich platform to reduce self-assembly complexity and both launch mass and volume.


\subsection{Sizing and Power Scales of Mid-Term HSA-SPS}

Similarly to the design described in Ref.~\cite{ownIACpaper}, mid-term technology advancement is expected to improve current sandwich performance parameters reported in Ref.~\cite{jaffepaper} by about $40\%$. This yields an area-specific mass of about $12$~kg/m$^2$ including all power transmission electronic elements, an effective sandwich height, including protective packaging, of about $9$~cm, and a sunlight-to-RF sandwich module efficiency of about $17\%$. In addition, the thermal properties of mid-term HSA-SPS are expected to allow three-sun concentration on the photovoltaic surface of the sandwich modules. 

\begin{figure}[h]
	\begin{center}
		\includegraphics[width=0.4\textwidth]{{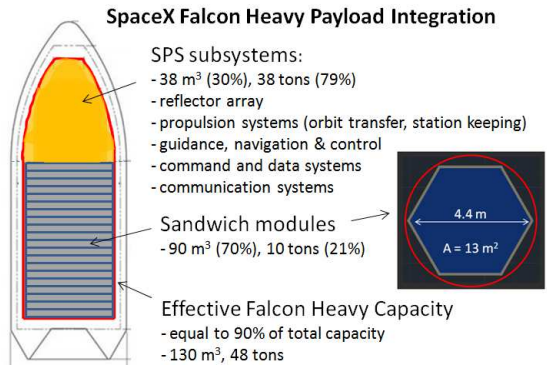}} 
		\caption{\label{fig:payloadintegration} 
		Illustrating payload integration scheme of a complete HSA-SPS in a Falcon Heavy vehicle.}  
	\end{center}
\end{figure}

As described above, the size of a complete HSA-SPS system is limited to the payload volume and payload mass of the SpaceX Falcon Heavy vehicle. Similar to the study in Ref.~\cite{ownIACpaper}, about $10\%$ are subtracted from the total payload volume of about $140$~m$^{3}$ for protective packaging against shock and vibration during launch. In the present study, however, sandwich modules are more stringently limited to occupy $70\%$ of the remaining volume or about $90$~m$^{3}$, to leave sufficient volume (about $38$~m$^{3}$) for the reflector array and all other HSA-SPS subsystems. This is assumed to improve the realistic value of the HSA-SPS design. Figure~\ref{fig:payloadintegration} shows a sketch of the payload integration of a complete HSA-SPS in the payload fairing of a Falcon Heavy vehicle.

Considering the expected mid-term effective sandwich module height of about $9$~cm, about $66$ hexagonal sandwich modules with base areas of about $13$~m$^{2}$ stacked along the vertical axis of the Falcon Heavy payload volume fulfill the volume constraints. Assuming that $90\%$ of the total, about $820$~m$^{2}$ power generating surface area of the sandwich modules are used to generate electricity at $3$-sun concentration with an incident energy density of about $1400$~W/m$^{2}$, a single Falcon Heavy can launch a HSA-SPS with a nameplate RF power generation capacity of about $520$~kW. 

Similarly to the studies in Ref.~\cite{ownIACpaper}, the sandwich platform of each HSA-SPS will self-assemble on orbit through e.g. spring-loaded interconnects between sandwich modules or other low-complexity technology options. As an illustration, Table~\ref{tab:midtermcomp} compares the specifications of sandwich platforms performing at current sandwich integration results with sandwich platforms performing along assumed mid-term technology advances.

	\begin{table}[t]
		\caption{Comparison between sandwich platforms consisting of sandwiches with currently achieved module parameters and expected mid-term improvements of sandwich parameters for utility-scale mid-term HSA-SPS. Resulting nameplate RF power levels for complete HSA-SPS from single Falcon Heavy launches are given as well. Current sandwich performance parameters are taken from Refs.~\cite{jaffepaper} and~\cite{perscorrJaffe}.}
\label{tab:midtermcomp}
	\centering
		\begin{tabular}{|c|c|c|} \hline
		          & Current Research & Mid-term HSA-SPS \\ \hline
		Mod. Area & $13$m$^{2}$        & $13$m$^{2}$ \\ \hline
		Eff. Mod. Height & $15$cm        & $9$cm \\ \hline
		Mod. Mass & $250$~kg           & $150$~kg \\ \hline
		Mod. Eff. & $12\%$            & $17\%$ \\ \hline
   \end{tabular} \\ \vspace{0.2cm}
Single Launch Sandwich Platform Performance \\
   \begin{tabular}{|c|c|c|} \hline
		Nr. Mod.  & $39$               & $66$ \\ \hline
		Tot. Mass & $10$~tons       & $10$~tons \\ \hline
		Tot. Area & $490$~m$^{2}$     & $820$~m$^{2}$\\ \hline
		RF Power  & $75$~kW (one-sun) & $520$~kW (three-sun) \\ \hline
						\end{tabular}
	\end{table}

Since the total mass of all sandwiches is about $10$~tons or about $21\%$ of the total payload mass capacity, the nameplate RF power level of HSA-SPS is entirely limited by the volume available in a Falcon Heavy launcher, or equivalently, by the effective height of the sandwich modules. The module number limit from the effective payload mass fraction of $70\%$ for sandwich modules is equal to about $180$ modules.


\subsection{HSA-SPS Subsystems and Reflector Array}

The remaining volume of $38$~m$^{3}$ and mass of about $38$~tons not occupied by sandwich modules is used for all remaining HSA-SPS subsystems. Similar to Ref.~\cite{ownIACpaper}, the reflector mass can be estimated using a reflector area mass density quoted in Ref.~\cite{IanRadicipaper} of $0.45$kg/m$^{2}$. Assuming that the reflector array will be at least three times as large as the area of all sandwich modules in an HSA-SPS to provide three-sun concentration, the total area will be about $2500$~m$^{2}$ (equal to a square area of $50$~m side length), corresponding to a total mass of about $1.1$~tons. Support structures and frames to ensure planarity are estimated to contribute about a similar amount, given carbon fiber solar sail boom research reporting mass per length densities as low as $0.06$~kg/m~\cite{Blocketal}. More elaborate reflector array architectures might be more suitable to maximize and homogenize the time full $3$-sun concentrated light across the power generation surface of the sandwich platform. Examples of such array concepts are given in Ref.~\cite{mankinsniac}. The mass of such arrays is expected to remain within the same order of magnitude as the estimates given above. On orbit, the reflectors can be deployed via inflation or roll-out technology.

\begin{figure}[h]
	\begin{center}
		\includegraphics[width=0.4\textwidth]{{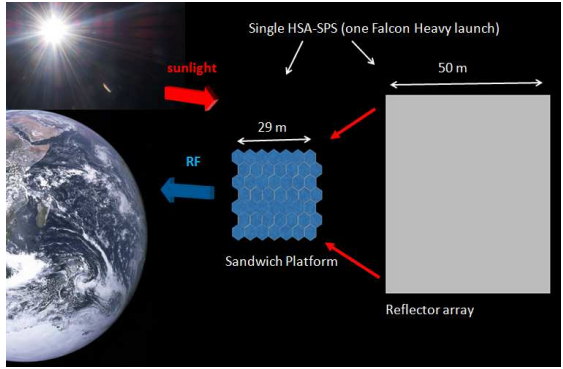}} 
		\caption{\label{fig:hsaspssinglelaunch} 
		A possible realization of a HSA-SPS system launched by a single Falcon Heavy vehicle. The hexagonal sandwich modules are arranged in a square shape with about $29$~m side length. The reflector array is reduced to symbolic features for simplicity and is of about $50$~m side length. The HSA-SPS is shown passing over the receiver site at local midnight. The view direction of the drawing is east along the SPS' orbit; the distance between the platform and the reflector array is chosen to facilitate display and not to scale.}  
	\end{center}
\end{figure}

Figure~\ref{fig:hsaspssinglelaunch} shows an illustration of a schematic realization of a HSA-SPS system, including the sandwich platform and the reflector array. The latter is reduced to be of symbolic character only for simplicity.

\subsection{Orbit Selection and Orbit Transfer}

As introduced in Ref.~\cite{ownIACpaper}, HSA-SPS are placed in geosynchronous Laplace plane orbits (GLPOs) where most orbital perturbations cancel, which allows removal of most of the propellant requirements for orbital station-keeping~\cite{Ianisdc13}. Since no in-space orbit transfer infrastructure is foreseen for HSA-SSP, each HSA-SPS is brought to GLPO after launch to LEO by on-board electric propulsion thrusters, similar to the design proposed in Ref.~\cite{ownIACpaper}: Electric propulsion systems with $I_{SP}=3000$~s can achieve the necessary $\Delta v$ for geostationary orbit (GEO) from LEO of about $4.7$~km/s using a spiral transfer orbit. Considering the payload mass constraints of a typical near-term LEO launch vehicle, e.g. the SpaceX Falcon Heavy launcher with a payload capacity of $53$~tons, a maximum payload net mass of about $45$~tons can be brought to GLPO with about $7.7$~tons of propellant. If low-cost argon is used as the propellant, the volume of the storage container for $7.7$ tons of argon can be estimated to about $7$~m$^{3}$. This leaves about $31$~tons and $31$~m$^{3}$ for all other satellite subsystems. 

\begin{figure}[h]
	\begin{center}
		\includegraphics[width=0.4\textwidth]{{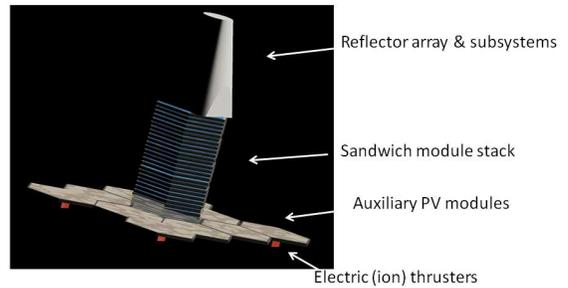}} 
		\caption{\label{fig:orbtrans} 
		A possible realization of orbital transfer from LEO to GEO for a complete HSA-SPS system. Electric propulsion thrusters are indicated by red elements on the bottom of the stack of sandwich modules. The black cone on the top represents the undeployed reflector array and other satellite subsystems. The auxiliary power generation panels supplying electricity to the thrusters are shown in brown-gray at the bottom of the stack of sandwich modules.}  
	\end{center}
\end{figure}

Analogously to Ref.~\cite{ownIACpaper}, the electric thrusters are powered by auxiliary photovoltaic (PV) modules. Assuming $10$ PV panels with similar active surface area as the sandwich module and about $20\%$ of efficiency, about $31$~kW of power can be provided. During the lengthy orbital transfer period, the propulsion photovoltaic modules are expected to suffer radiation and micrometeoroid damage and reduced efficiency. However, they can be used as additional energy sources for $3$-axis-stabilization to maintain optimal positioning of the deployed HD-SPS in GLPO. Figure~\ref{fig:orbtrans} shows a possible realization of the orbital transfer setup of a complete HSA-SPS unit which remains undeployed for transfer to minimize its cross section versus space weather and orbital debris effects and impacts.

\section{Exemplary HSA-SPS Constellation for Utility-Scale Baseload Power}

\subsection{HSA-SPS GLPO Constellation Architecture}

In order to achieve $1$~GW of RF power on orbit, about $1900$ HSA-SPS have to be launched and transferred to GLPO with the same number of Falcon Heavy launches. For simplicity, this Section will discuss a continuous, circular constellation of HSA-SPS where all SPS are physically connected to each other with simple mechanical locking-mechanisms. The so-formed circular constellation has a diameter of about $1400$~m. Each HSA-SPS still operates independently, such that faulty SPS can quickly be removed from the constellation and exchanged.

The free-flying reflector array of a utility-scale HSA-SPS constellation can be formed by unconnected or connected HSA-SPS reflector elements. The distance between the circular sandwich platform and the reflector array in radial direction form Earth can be optimized to achieve best light concentration throughout the entire orbit. As a loss inherent to the design principle of this architecture, the umbra and penumbra shadows of the sandwich platform will move across the reflector array once every day around the midnight hours. If the distance between the sandwich platform and reflector array is kept small, the shadow of the sandwich platform will project nearly by parallel lines onto the reflector array and occult nearly exclusively an area equal to that of the sandwich platform. Therefore, during these few hours per day the level of sunlight concentration delivered by the reflector array onto the photovoltaic surface is reduced from $3$ to a minimum of $2$. Integrated over a full day, a reduction of at most $17\%$ in total RF energy produced is caused compared to continuous $3$-sun concentration. Considering the cost savings because of the independence of yet-to-be-developed in-space assembly and orbital transfer infrastructure, this drawback seems to be acceptable. In addition, lower RF power will only be delivered during midnight hours where power demand is expected to be at its daily minimum.

To minimize the sandwich platform shadow size on the reflector, small in-plane displacements between the platform and the reflectors, e.g. $3$~km, should be favored. In order to maintain the same orbital period between the sandwich platform and the reflector array displaced by $3$~km, the latter has to be kept on its orbit with an orbital velocity increase of $0.33$~m/s. Assuming $2$~tons of mass for each HSA-SPS reflector, a continuous force of about $0.13$~N has to be applied to achieve the same orbital period as the sandwich platform without leaving the reflector's circular orbit around Earth. The NSTAR ion thruster of NASA has achieved maximum thrust of about $0.091$~N at a power level of $2.3$~kW and maximum specific impulse of $3120$~s~\cite{nasanextpaperIAC}. This corresponds to a propellant mass needed of about $2.8$~tons for a $30$-year duration mission. The required power can be provided by the auxiliary photovoltaic models used in orbit transfer. It is therefore assumed that the conditions to fly the reflector array at the same orbital period as the sandwich platform safely already with current technology, while reserving about $3$~tons of payload mass and $3$~m$^{3}$ of volume for the required propellant.


\subsection{Opportunities for Thermal Management Improvements}

For improved thermal management, the locking mechanisms between single HSA-SPS can be adjusted to leave e.g. $50$~cm of empty space between all HSA-SPS. In these gaps, additional radiator material could be placed for additional heat rejection (e.g. protective material layer around the payload stack during orbit transfer), however without obstructing the paths of incoming sunlight and outgoing RF power. The inter-SPS distance has to be kept small in order to limit RF sidelobe losses due to the 'thinned array curse' of spatially separated radiation sources, c.f. Ref.~\cite{thinnedarraycurse}.  

\subsection{HSA-SPS Ground Systems}

Similar to the design proposed in Ref.~\cite{ownIACpaper}, a rectenna is used to convert intercepted RF power to electricity. A corresponding ground station will also involve a pilot signal generation facility to enable retrodirective RF beam pointing and phase control and command and controls center for operating the power satellites and to monitor the RF beams. For a utility-scale HSA-SPS constellation providing $1$~GW of RF power, using the relation

\begin{equation}
\tau = \frac{\sqrt{A_{t}A_{r}}}{\lambda D}
\label{eq:taueq}
\end{equation}
yields a rectenna diameter of about $7.9$~km to intercept more than $95\%$ ($\tau~2$) of the generated RF power~\cite{jaffepaper}. The rectenna site could be constructed similarly to existing large RF facilities, such as the HAARP array in Alaska, USA~\cite{haarpwebpage}. This array involves an RF power generating capacity on the MW scale (most likely less will be needed to operate a pilot signal), an antenna phased array of size $300 \times 370$~m and beam safety systems such as aircraft alert radar.

In the context of the sandwich platform shadow on the reflector plane described above, continuous baseload power at the $66\%$ level of nameplate RF power can still be supplied to the power grid by using the excess power delivered during the day in e.g. hydrogen fuel production or saltwater desalination as a variable load as described in Ref.~\cite{sneadpaper}.


\section{High-Level Cost Estimates for Utility-Scale HSA-SSP Constellation}
\label{sec:leodemocost}

Similarly to Ref.~\cite{ownIACpaper}, a learning curve-based order of magnitude cost analysis is performed, here for a utility-scale constellation of about $1900$ HSA-SPS systems providing about $1$~GW RF power. Again, the cost for hardware and production is separated from launch and research and development, test and evaluation (RDT\&E) cost. An initial sandwich production and hardware cost of $100,000$~USD/kg is assumed, yielding a theoretical cost of the first sandwich module of about $15$~M~USD. RDT\&E costs are estimated as $5.5$ times the first module cost~\cite{IanRadicipaper} and yield about $83$~M~USD for the sandwich modules. Also analogously to Ref.~\cite{ownIACpaper}, a drop in production unit cost by $34\%$ for every doubling of the number of units produced is assumed~\cite{mankinsniac}, where the total cost is determined from integrating the cost curve up to the total module number. The drop in launch cost is again parametrized more conservatively with a cost drop of $27\%$ at doubling of launches since launch cost might drop more slowly than SPS production cost. Following the methods outlined above, the about $130,000$ single sandwich modules required for a $1$~GW RF power HSA-SPS constellation can be produced for about $4.2$~B~USD.

In this cost analysis, all remaining subsystems are divided into the following four categories: reflector array, self-deployment and payload integration, orbit transfer, and remaining generic subsystems such as guidance, navigation and control, command and data systems and communication systems. These categories are in the following treated with in similar ways. It is assumed that every categories of subsystems can be produced for $10\%$ each of the cost of the sandwich modules per HSA-SPS (since much less subsystem units are needed per HSA-SPS compared to sandwich modules). This adds about $1.7$~B~USD to the total cost. To estimate RDT$\&$E cost, in the presented analysis it is assumed that the first unit of each of the four categories of subsystems can be developed for the same cost as the first sandwich module. This yields a total RDT$\&$E cost of about $330$~M~USD for all four categories. 

Recently, SpaceX updated their launch cost for a launch of a Falcon Heavy to LEO to $77$~M~USD~\cite{spacex}. Applying the more moderate learning curve approach to cost development of launching about $1900$ HSA-SPS to LEO, a total launch cost of about $8.7$~B~USD is obtained. 

Following Ref.~\cite{mankinsniac}, unit area cost for the rectenna is estimated by $10$~USD/m$^{2}$. For the $7.9$~km diameter rectenna discussed in this paper, the cost for the ground rectenna amounts to about $490$~M~USD. A recurring annual program/operations cost of $5.5$~M~USD/year is also assumed in Ref.~\cite{mankinsniac}, which is neglected in this analysis. Ground-based RDT\&E cost for the rectenna is estimated as three times the cost of a $1$~km$^{2}$ rectenna, amounting to $30$~M~USD.

A summary of the estimated cost contributions for development, production and launch for a $1$~GW RF power HSA-SPS constellation in GLPO is given in Table~\ref{tab:hsaspscost}, in addition to costs for a rectenna ground station. A total cost of about $15$~B~USD is estimated for the entire constellation and ground systems. The numbers quoted in this Section are only order-of-magnitude estimations and do not claim to be complete or comprehensive.

	\begin{table}[h]
		\caption{Summary of estimated cost contributions for RDT$\&$E, production, launch and ground systems of a $1$~GW RF power HSA-SPS constellation in GLPO.}
\label{tab:hsaspscost}
	\centering
		\begin{tabular}{|c|c|} \hline
        {\bf Item}      & {\bf Est. Cost + RDT\&E Cost} \\ \hline
Sandwich Production     & $4.2$~B~USD + $83$~M~USD \\ \hline
Other Satellite Systems & $1.7$~B~USD + $330$~M~USD\\ \hline
Launch                  & $8.7$~B~USD \\ \hline
Rectenna                & $490$ + $30$~M~USD \\ \hline
{\bf Total }            & $15$~B~USD \\ \hline
				\end{tabular}
	\end{table}

In a similar reasoning as in Ref.~\cite{ownIACpaper}, the estimated development, production and launch cost of a $1$~GW RF power HSA-SPS constellation would amortize within about $13$ years, if the rectenna could be located in New England and power could be sold to utility companies for $14$~cents/kWh, so cheaper than the average retail price for all end-use sectors~\cite{eia.gov}. Again, earlier break-even could be reached if power is being sold already before all HSA-SPS are delivered to GLPO.



\section{Advantages of the HSA-SSP Design}

The main design advantages of HSA-SSS are shared with the design proposed in Ref.~\cite{ownIACpaper}: independence of costly and yet-to-be developed in-space assembly and orbital transfer infrastructure; sale of utility-scale power to utility companies within probably less than $10$ years of production start (assuming $190$ Falcon Heavy launches per year); reduction of development time due to limited size of complete HSA-SPS systems. Also, damaged HSA-SPS can be removed from the constellation, repaired in a repair orbit and then re-inserted into the constellation, while new HSA-SPS can be added to the constellation at any time via Falcon Heavy launches.

Critically, the HSA-SPS design at the present level of study does not pose inherent, with current or mid-term technology insurmountable orbital dynamic or RF beam forming difficulties, in contrast to the design shown in Ref.~\cite{ownIACpaper}. The HSA-SPS design is fairly straight forward in terms of only consisting of two large-scale elements on the architecture level, which are the sandwich platform and the reflector array. Distance between these two elements can be chosen such that the difference in orbital velocity and period can be remedied via electric thrusters with a moderate propellant requirement. Placement of the elements in GLPO should largely reduce the two necessary station-keeping efforts for pointing the Rf transmission surfaces at Earth at all times and directing concentrated sunlight on the photovoltaic surfaces from the reflector arrays. Additional orbital dynamic studies should be performed to verify the validity and low technological complexity of the design, and that system lifetimes of the order of $30$~years are realistic.

Thermal performance of HSA-SPS can be further enhanced, as mentioned above, by introducing gaps between adjacent single HSA-SPS and utilizing the gaps for radiator material to decrease the operating temperature of the sandwich module. In addition, similar to the design shown in Ref.~\cite{ownIACpaper}, HSA-SPS are sufficiently small such that any sandwich module is fairly close to potential radiators on the outsides of the SPS. This would enable fairly efficient transport of excess heat from any sandwich module via heat pipes to the outside radiators, which could allow higher concentration factors on the PV modules. Another possible opportunity for the HSA-SPS design is that electrical power could be shared among neighboring sandwich modules or among modules in the entire HSA-SPS.


\section{The Path Forward: Considerations for Near-Term Sandwich Module Development}

In order to optimize economics of SPS launch, it is generally accepted that area-specific mass should be decreased and efficiency increased. However, in the context of pre-fabricated and self-deploying SPS designs, sandwich volume, or effective module height for a given base surface area, is identified as a critical parameter. To most economically use e.g. the SpaceX Falcon Heavy vehicle capacities, an average effective payload density of about $0.37$~tons/m$^{3}$ has to be achieved, if SSP remains the only large-scale buyer on the launch market. For an assumed mid-term $10$~kg/m$^2$ area-specific mass of sandwich modules, a corresponding effective module height of about $2.7$~cm is necessary to attain the optimal module mass density, and thus to maximize the number of modules transported per launch. Considering that the mass of about $180$ sandwich modules with parameters given in Table~\ref{tab:midtermcomp} could be launched by a single Falcon Heavy vehicle, but the volume of only $66$ modules, module height can be regarded as a more limiting aspect in sandwich payload integration than area-specific mass at this time.

If in-space assembly infrastructure is not developed in the next decades, dedicated R$\&$D effort should be focused in the near-term on reducing the effective height of SPS sandwich modules, e.g. to less than $5$~cm. This could reduce SPS launch costs by a factor $2$ or more. A possible solution to this problem could be attempting to integrate PMAD and RF power electronics into a heat-distributing substrate attached to high-efficiency thin-film PV, combined with thin RF antennas. Planarity could be supported by, e.g., an aluminum mesh. The thermal properties of such a system remain to be investigated.


\section{Conclusion}

An update of the SSP design presented in Ref.~\cite{ownIACpaper} results in the formulation of a new SSP concept, HSA-SSP. The design is scalable to utility-scale power and does not rely on in-space assembly or transportation infrastructure. Extrapolations of current technology suggests a realistic design for mid-term SPS deployment. A cost estimation analysis for a $1$~GW RF power HSA-SPS constellation yields a total cost of about $15$~B~USD, utilizing a learning curve approach similar to the one described in Ref.~\cite{mankinsniac}. The advantages of the presented design will be solidified and extended in further research.

\section*{Acknowledgment}
I would like to thank Mr. Ian McNally from the University of Glasgow, Scotland, UK and Mr. Paul Jaffe from the Naval Research Laboratory, USA for helpful discussions and suggestions.





\begin{thebibliography}{99}

\bibitem{glaser}
P.~E.~Glaser, 
Science, {\bf 162} $3856$, $857$ ($1968$).

\bibitem{iaastudy2011}
For a comprehensive study about technical SSP feasibility, see \\
J.~C.~Mankins (editor), Cosmic Study/Position Paper from the International Academy of Astronautics ($2011$).

\bibitem{mankinsniac}
J.~C.~Mankins, 
NIAC Phase $1$ Final Report ($2012$).

\bibitem{leopoldpaper}
L.~Summerer and L.~Jacques, 
Proceedings to International Astronautical Congress $2011$, IAC-$11$.C$3$.$1$.$3$.

\bibitem{jaffepaper}
P.~Jaffe and J. McSpadden, 
Proceedings of the IEEE {\bf 101}, $1424$ ($2013$).

\bibitem{ieareport}
International Energy Agency (IEA), World Energy Outlook $2012$, Executive Summary.

\bibitem{ipcc4threportwg1}
S.~Solomon {\it et al.}, 
Intergovernmental Panel on Climate Change (IPCC) Assessment Report $4$, Working Group $1$ ($2007$), 
Cambridge University Press.

\bibitem{ownIACpaper}
M.~Leitgab, to be published in
Proceedings to International Astronautical Congress $2013$, IAC-$13$.C$3$.$1$.$9$.x$19763$.

\bibitem{spacex}
See \url{http://www.spacex.com}, last accessed $09/27/2013$.
	 
\bibitem{perscorrJaffe}
Personal correspondence with Mr. P. Jaffe, NRL, USA.

\bibitem{IanRadicipaper}
I.~McNally, M.~Ceriotti and G.~Radice,
Proceedings to International Astronautical Congress $2012$, IAC-$12$-C$3$.$1$.$9$.

\bibitem{Blocketal}
J.~Block, M.~Straubel and M.~Wiedemann, Acta Astronautica {\bf 68}, $984$ ($2011$).


\bibitem{Ianisdc13}
I.~McNally, D.~Scheeres, G.~Radice, International Space Development Conference 2013, San Diego, CA, USA.

\bibitem{nasanextpaperIAC}
G.~R.~Schmidt, M.~J.~Patterson, and S.~W.~Benson, 
Proceedings to International Astronautical Congress $2008$, IAC-$08$-C$4$.$4$.$2$.

\bibitem{thinnedarraycurse}
See \url{http://en.wikipedia.org/wiki/Thinned-array_curse}, accessed $09/27/2013$.

\bibitem{haarpwebpage}
See \url{http://en.wikipedia.org/wiki/High_Frequency_Active_Auroral_Research_Program}, accessed $09/27/2013$.

\bibitem{sneadpaper}
J.~M.~Snead, White Paper, 
\url{http://mikesnead.net/resources/spacefaring/white_paper_the_end_of_easy_energy_and_what_to_do_about_it.pdf}, accessed $09$/$27$/$13$.

\bibitem{eia.gov}
See \url{http://www.eia.gov/electricity/monthly/epm_table_grapher.cfm?t=epmt_5_6_a}, accessed $09/27/2013$.

\end{thebibliography}
%
%

\end{document}